# The Detector System Design for the Grating-based Phase Contrast Imaging CT Prototype

Lian Chen, Rongqi Sun, and Ge Jin

*Abstract–In the hard X rays domain, the phase shift of the wave passing through the soft materials like tissues is typically three orders of magnitude larger than the absorption. Therefore the phase-sensitive X-ray imaging methods can obtain substantially increased contrast over conventional absorption-based imaging. In this paper, we present a detector system design for the computed tomography(CT) prototype which based on the grating-based phase contrast imaging(GBPCI) method. This system consists of 43 Hamamatsu silicon photodiode S12058(X) modules including 16512 pixels. Each module can provide 384(16\*24) pixels, which the minimum pixel size is 0.75mm\*1mm. The front-end readout electronics(FEE) of each module complete the analog-to-digital conversion, the digital processing, and the data transmission. All digitized data from 43 detector module are transmitted to the control system through the data collection boards. The test result shows that the efficient resolution of the detector system is up to 14.6bit. For all pixels, the integral nonlinearity is no more than 0.1%.*

## I. Introduction

X-ray radiographic absorption imaging is a standard tool in the medical diagnostics reasearch area. However, the weak absorbing structures, such as biological tissue, cannot be imaged with a high sensitivity under the X-ray dosage limit. Theoretically, the interaction cross section of X-ray phase shift is usually much greater than the absorption cross section for soft tissues. Therefore, recording the X-ray phase shift can obtain substantially increased contrast. In the past years, various phase-sensitive X-ray imaging mehtods were developed, such as refraction-based method, propagation-based method, analyzer crystal-based method[1-3]. Nevertheless, the application of these methods in medical diagnostics is very slow, because a much higher quality X-ray source is required than that of conventional laboratory or hospital X-ray sources.

In 2006, a three transmission gratings phase-contrast image method with conventional X-ray tubes is reported[4]. The G0 granting creates an array of individually coherent, but mutually incoherent sources. Through the combination of gratings G1 and G2, the small angular deviation changes the locally transmitted intensity. The method can simultaneously obtains the separate absorption information and phase-contrast images, which can achieve a large field of view with low requirement on time and spatial coherence of the light source. Therefore, the GBPCI method is considered as a potential imaging method for clinical applications. A CT prototype based on GBPCI is developed by National Synchrotron Radiation Laboratory (NSTL).

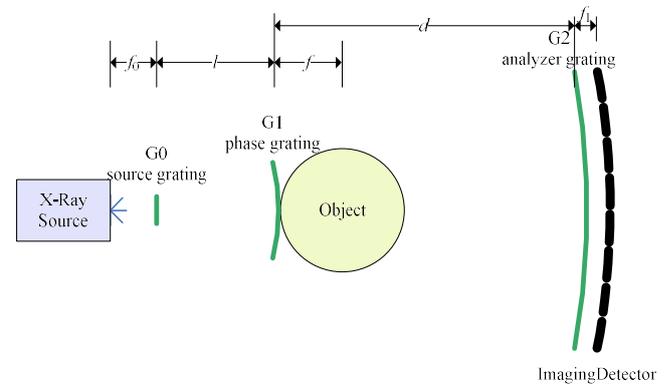

FIG. 1. The schematic diagram of three grating phase-contrast image method

In imaging system, the effects of light source and the detector on the system spatial resolution can be described as follows:

$$\delta = \sqrt{\delta_w^2 + \delta_{p_d}^2}$$
$$\delta w = w/2M = \left(s\frac{l}{f_0+l}\frac{d-f}{l+f}\right)\Big/2M \quad (1)$$
$$\delta p_d = p_d/2M$$
$$M = (d-f/2)/(l+f/2)$$

which $s$ is light source size, $p_d$ is the detector pixel size, $M$ is the optical magnification.

In this CT prototype, the distance $d$ between the G1 grating and the G2 grating and the distance $l$ between the G0 grating and the G1 grating are 1200mm and 300mm respectively, which to balance the requirements of measurement sensitivity and X-ray intensity. In order to achieve the spatial resolution of 0.2mm and cover the field of view of $\varphi$200mm, the pixel size of the imaging detector should be less than 0.75mm and the horizontal effective width of the imaging detector should reach about 700mm. A prototype detector system is designed to satisfy these requirements.

Manuscript received June 13, 2018. This work was supported by National Key Scientific Instrument and Equipment Development Project under Grant No. ZDYZ2014-2.

Lian Chen, Rongqi Sun, and Ge Jin are with State Key Laboratory of Particle Detection and Electronics, University of Science and Technology of China, Hefei, Anhui 230026 China (phone:+86-0551-63607152,e-mail:chenlian@ustc.edu.cn, srq@mail.ustc.edu.cn, goldjin@ustc.edu.cn)

## II. SYSTEM DESIGN

The detector system consists of 43 Hamamatsu silicon photodiode S12058(X) modules to achieve 200mm*200mm measurement field. Each S12058(X) module can provide 384(16*24) pixels, which the minimum pixel size is 0.75mm*1mm. The FEE board is direct connected to each S12058(X) module. The structure of FEE is shown in Fig.2.

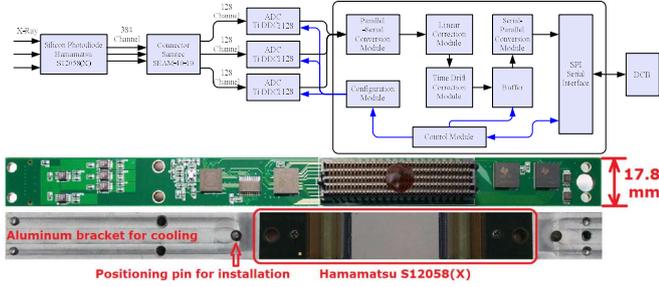

FIG. 2. The schematic diagram (above) and the prototype board (below) of the FEE

The S12058(X) module receives the X-rays and generates 384 current signals whose electric charge are related to the intensity of X-rays. These current signals are transmitted to the FEE via a high density connector (SAMTEC SEAM-40-10). Three Ti DDC1128 analog-to-digital converts(ADCs) are used to complete both current-to-voltage and A/D conversion. The serial digital data transmit into the FPGA and convert to the parallel data.

After the linear correction and time drift correction, the image data are sent to the data collection board(DCB) through the data buffer.

Consider to the system installation, the width size of FEE must smaller than the detector. The size of S12058(x) module is 18mm*100mm, thus the width of FEE is limited to 17.8mm. Due to the large number of pixels, the peak power consumption of each FEE can reach to 3W. An aluminum bracket with 10mm thickness is used to export the heat which generated by each FEE, and an air-cooled system is setting to cool the whole detector system.

## III. TEST RESULT

A noise test is performed to evaluate the efficient resolution of the detector system. The S12058(X) module is set to test mode, then all the pixel inputs are connected to the GND. The equicalent noise of each pixel is calculated by the Full-Width-Half-Maximum of the Gaussian distribution fitted to the output amplitude, as is shown in Fig.3. The average noise of the detector system is about 6.2fC. The efficient resolution of the detector system can be calculated as : N-log($\sigma$) = 14.6bit.

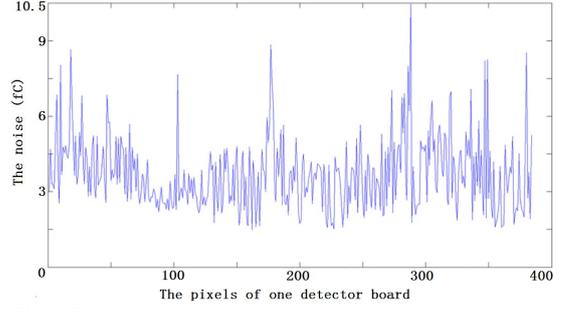

FIG.3 The equicalent noise of 384 pixels on one detector board

For the diversity of each detector module and the difference of each FEE channel, the gain uniformity test is performed. As is shown in Fig.4, the gain where located on the edge of the S12058(X) module is evidently lower than other pixels which caused by the edge effect. An air scan calibration method is used to correct the consistency.

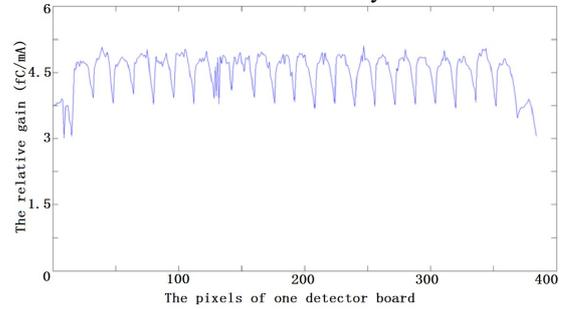

FIG.4 The relative gain of 384 pixels on one detector board

The CTR2150 X-ray source is used to perform the linearity test. By stepping up the output current of the source, the incoming charge quanntity of each pixel is synchronous increased linearity. The integral nonlinearity(INL) of the detector board can be obtained with simple calculation. As is shown in Fig.5. The result shows that the detector system has a good linear response in the signal dynamic range, and the INL of 384 pixels on one detector board is less than 0.1%.

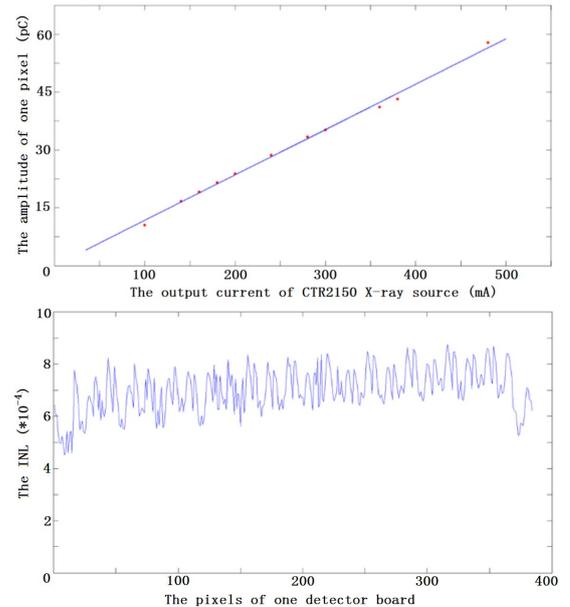

FIG.5 The linearity of one pixel(above) and the INL of 384 pixels on one detector board(bleow)

A desktop experiment platform is set up to measuring the spatial resolution. The simplified detector system includes 5 detector boards to accommodate the size of the desktop platform. A lead resolution board is used as the test sample. The test result is shown in Fig.6. The spatial resolution of the detector system can be up to 0.38mm with a 17% contrast after air scan calibration.

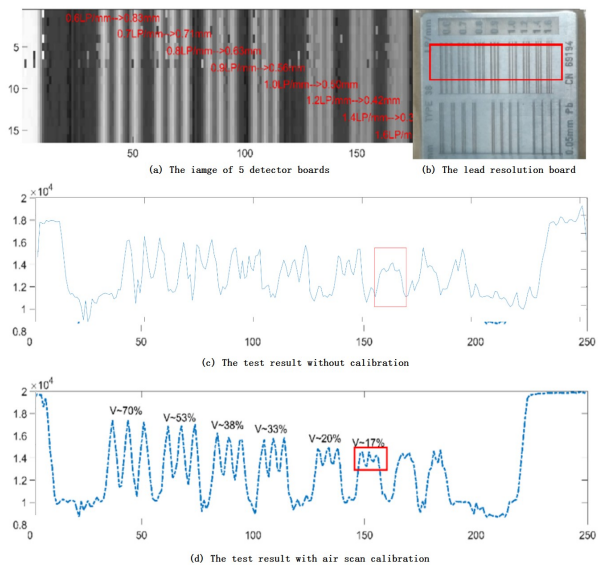

FIG.6 The spatial resolution test results of desktop experiment platform

## IV. CONCLUSION

In this paper, we present a detector system for the grating-based phase contrast imaging CT prototype. The system contains 43 detector boards including 16512 pixels. Test result shows that the average efficient resolution is about 14.6 bits. The INL of all pixels is less than 0.1%. The air scan algorithm is used to calibrate the gain non-uniformity. The spatial resolution of the system can be up to 0.38mm with a 17% contrast which close to the theoretical value 0.36mm for the principle experiment.

REFERENCES

[1] A. Momose, et al. "Phase-contrast X-ray computed tomography for observing biological soft tissues," Nature Med. 2, 473–475 (1996).
[2] T. J. Davis, et al. "Phase-contrast imaging of weakly absorbing materials using hard X-rays," Nature (London) 373, 595–598 (1995).
[3] P. Cloetens, et al. "Holotomography: Quantitative phase tomography with micrometer resolution using hard synchrotron radiation x rays," Appl. Phys. Lett. 75, 2912–2914 (1999).
[4] Pfeiffer F, et al. "Phase retrieval and differential phase-contrast imaging with low-brilliance X-ray sources." (4)2,258-261(2006).